\documentclass[showpacs,showkeys,floatfix]{revtex4}
\usepackage[dvips]{graphicx}
\usepackage{epsfig}

\usepackage[cp1251]{inputenc}
\usepackage[english]{babel}

\begin{document}

\title{ Pyrophosphate Groups in Liquid Crystalline Phases  of the DNA }

\author{V.L. Golo$^1$}
\email{voislav.golo@gmail.com }
\author{E.I.Kats$^2$}
\email{kats@ill.fr}
\author{S.A. Kuznetsova$^3$}
\email{svetlana@belozersky.msu.ru }
\author{Yu.S. Volkov$^1$}
\email{yu.volkov@gmail.com}

\affiliation{$^1$ Department of Mechanics and Mathematics,
                  the Lomonosov Moscow State University,  Moscow 119 992 GSP-2, Russia   }
\affiliation{$^2$ Institute Laue-Langevin     Grenoble, France and L.D.Landau Institute for Theoretical Physics, Moscow, Russia}
\affiliation{$^3$ Department of
                  Chemistry,  the Lomonosov Moscow State University,  Moscow 119 992 GSP-2, Russia   }

\date{August 3,  2008}

\begin{abstract}
We study electrostatic interaction between molecules of the DNA in which a number of phosphate groups
of the sugar-phosphate backbone are exchanged for the pyrophosphate ones.  We employ a model in which
the DNA is considered as a one-dimensional lattice of dipoles and charges corresponding to base pairs
and (pyro)phosphate groups, respectively. The interaction between molecules of the DNA is described by
a pair potential $U$ of electrostatic forces between the two sets of dipoles and charges belonging to
respective  lattices describing the molecules. Minima of potential $ U$ indicate orientational
ordering of the molecules and thus liquid crystalline phases of the DNA.  We use numerical methods for
finding  the set of  minima in conjunction with symmetries verified by potential $U$  . The symmetries
form a noncommutative group of 8-th order, ${\cal S}$. Using the group ${\cal S}$ we suggest a
classification of liquid crystalline phases of the DNA, which allows of  several cholesteric phases,
that is polymorphism. Pyrophosphate forms of the DNA could clarify the part played by charges in its
liquid crystalline  phases, and  make for experimental research, important for nano-technological and
bio-medical applications.

\end{abstract}


\maketitle

\section{Introduction}

According to the familiar legend the discovery of the cholesteric phase of the DNA was due to a happy
chance that occurred to C.Robinson, who was sharp enough to see it, \cite{robinson}.   The story
resembles that of Fleming's discovering the penicilin. Since then cholesteric phases of the DNA have
been the subject of numerous  experimental and theoretical investigations owing to their variety and
regularity, \cite{livolant1}. It has been established that the formation of the phases depends on
chemical properties of an ambient solution and ions, the ingenious experimental technique has been
worked out, \cite{livolant1}, \cite{livolant2}, and liquid crystalline phases of the DNA have become
instrumental in studying the DNA itself.  Another important development  began surfacing in chemical
physics of nucleic acids early in the 90-th. The  group led by Z.A.Shabarova at the Lomonosov Moscow
University, \cite{s1} --- \cite{s5} succeeded in synthesizing  the DNA in which some inter-nucleotide
phosphate groups are exchanged for the pyro-phosphate ones, and thus considerably extended the field
of research, providing new insights into the chemistry of nucleic acids, as well as new possible
bio-medical applications. In this paper we aim at making it clear that pyrophosphate forms of the DNA
could be helpful in studying liquid crystalline phases of the DNA.

Theoretical work on the physics of liquid crystalline phases dates from the seminal paper by Onsager,
\cite{onsager}, which relies on the  picture of hard rods representing molecules in solvent.
Applications of the model require the use of phenomenological constants and theoretical assumptions,
difficult to verify in specific situations. Cholesteric phases need even more careful investigating
owing to their chirality. In a series of papers Kornyshev, Leikin, and their
collaborators\cite{lk_model}, \cite{lk1} - \cite{lk3}, \cite{kim}, \cite{osipov}, put forward the
theory of cholesteric liquid crystalline phases of the DNA that relies on the helical distribution of
charges of the DNA. Within the framework of this theory one can employ various approaches and
approximations  and investigate specific conformations. Generally, a molecule of the DNA is considered
as a charged rod or cylinder, the charge being distributed continuously over the surface of the rod,
complying with the helical symmetry, theoretical tools employed being of analytical character. In the
present paper we use a discrete approximation for the charge's distribution and rely a computer
simulation for finding molecular conformations.  It should be noted that the distribution of charge in
the DNA molecule is essentially discrete being caused by (1) the dipole moments of the base pairs, (2)
the charges of the phosphate groups, (3) counterions which are not uniformly distributed round the DNA
molecule.  The electrostatic interaction between two DNA molecules is due to this essentially
non-uniform distribution of charges. Our approach, still remaining within the framework of papers
\cite{lk_model}, \cite{lk1}, provides new details of the phenomenon.  It is worth noting that we aim
only at a qualitative  description, which could be useful for explaining experimental data.

Since the current theory considers electrostatic interaction as a cause for the formation of liquid
crystalline phases the DNA , it is interesting to investigate opportunities that can be provided by
the use of the DNA containing a number of pyrophosphate groups, PP-forms, instead of the usual
phosphate ones, P-forms, see Figure 1, so as to have  a means of changing the charge conformation of
the molecule. It is important that synthetic forms of the DNA can contain PP-groups  in the duplex of
the DNA instead of the usual phosphate ones in such a way that the structures of the pyro-modified and
usual phosphate molecules, remain rather close,\cite{s5}, the inter-nucleotide distance, the stacking,
and the Watson-Crick interaction suffering no change.

Synthetic forms of the DNA  are instrumental in the study of fundamental problems in molecular
biology, biochemistry, medicine,  ferments' activity in nucleotide exchange, protein-nucleic acids
complexes, structural functioning of biopolymers, and regulation of the genetic expression. The
modification of inter-nucleotide groups is of particular importance owing to its preserving the
ability of molecules of the DNA to penetrate cell membranes and regulate gene expression, while
retaining the basic function of the DNA, that is  to interact with the complimentary sequences of
nucleotides. It is possible to synthesize the DNA so as to have  the exchanged pyrophosphate groups
located at prescribed sites of the sugar-phosphate spine, each pyrophosphate group bringing forth an
additional negative charge.

Minima of the potential {\cal U} for pair-interaction of molecules of the DNA should correspond to
orientational ordering of the molecules in solvent and therefore  liquid,  or solid, crystalline
phases. Special means are required to find the  minima of $ U$.  At this point the symmetry of {\cal
U} provides valuable information. As was found in the previous paper \cite{gkv}, $ U$ is invariant
under the action of a group of discrete transformations, and therefore its minima form a set having
the same symmetry.  The circumstance reduces the amount of numerical work, which is quite large.  But
we feel that the symmetry of the pair-interaction $ U$ is by far of more general importance for
understanding the physics of liquid crystalline phases of the DNA than one can infer from its
numerical applications.

\section{Preliminaries. }

We shall recall certain basic facts of the DNA. A molecule of the DNA can attain several hundred $\mu
m$ in length. If we neglect details that have a size of one thousand \AA, or more, we can visualize
it as a soft shapeless line and conclude that on this scale it behaves like an ordinary polymer. In
contrast, looking at its smaller segments, of one hundred \AA or less, we observe that it tends to
be straight.  Thus, borrowing a comparison from everyday life, we may say that a molecule of the DNA
looks like a piece of steel wire whose long segments are flexible and the short ones are stiff. The
elastic properties of the DNA are intimately related to its being a double helix. The latter imposes
severe constraints on deformations which can be effected without destroying the molecule and to a
large extent determines its mechanical properties. In fact, the two strands comprising the molecule of
DNA have just small bending rigidities, just as usual polymers. But the formation of the two-stranded
structure drastically changes the DNA by making it both stiff and capable of forming sophisticated
spatial shapes.

As was mentioned above, the double helix of DNA consists of long chains, or strands, which have the
backbones composed of sugar and phosphate residues, and special chemicals, bases, keeping the two
strands together (the structure is illustrated in Figure 2). The fundamental building blocks of the
strands are nucleotides, joined to each other in polynucleotide chains. The nucleotide consists of a
phosphate joined to a sugar (2'-deoxyribose), to which a base is attached. The sugar and base alone
are called a nucleoside. The chains, or strands, of the DNA wind round each other in a spiral forming
a double helix, the bases being arranged in pairs: adenine - thymine (AT), guanine - cytosine (GC), so
that the sequence of bases in one strand determines the complimentary sequence of bases in the other
and constitutes the genetic code stored by the molecule of  DNA. There are several forms of the DNA,
denoted by A,B, and Z. The most common one in nature, is the so-called B-form. One turn of the helix
of the B-form, corresponds approximately to $10.5$ base-pairs, and the distance between adjacent pairs
of bases is approximately $3.4$ \AA. In real life there are considerable deviations from the
canonical B-form of the DNA. Therefore, there is a need for a special nomenclature for describing its
conformations (see \cite{calladine} for the details),and generally a considerable set of parameters is
required. It is worth noting that the deviations from the canonical form are by no means small, and
may have a size of tens of degrees.

Synthetic analogues of nucleic acids (NA) containing modified internucleotide groups are useful for
solving various problems of molecular biology, biotechnology, and medicine.  Shabarova et al,
\cite{s1}, \cite{s2} ,developed a novel type of modified DNA duplexes containing pyrophosphate ( PP)
and substituted pyrophosphate ( SPP) internucleotide groups at the definite position of the
sugar-phosphate backbone \cite{s1} - \cite{s5} (see Figure 1).  The PP-group bears one additional
negative charge in comparison with a natural internucleotide group; SPP group contains no additional
charges. The introduction of PP-groups into DNA leads to an increase of total negative charge of a
molecule of the DNA. The study of oligonucleotide duplex containing a PP and SPP groups has revealed
that stacking and Watson-Crick interactions  are not significantly affected. By flipping out of the
disubstituted phosphate these groups fit into the helix structure without elongation of the
internucleotide distance. The analysis of helical parameters of base-pairs , internucleotide
distances, and overall global structure, reveals a close similarity of the initial and modified
duplexes.

The location of PP-groups of the DNA are to verify certain conditions:
\begin{enumerate}
    \item their total number does not exceed 10 \% of the total number of phosphates; \\
    \item they are not allowed to be located opposite each other; \\
    \item they are not allowed to occupy the ends of a molecule; \\
    \item two adjacent pyrophosphate groups are to be separated by at least 10 phosphate ones.
\end{enumerate}

\begin{figure}[htb]
    \centering
        \includegraphics[width=0.5\textwidth]{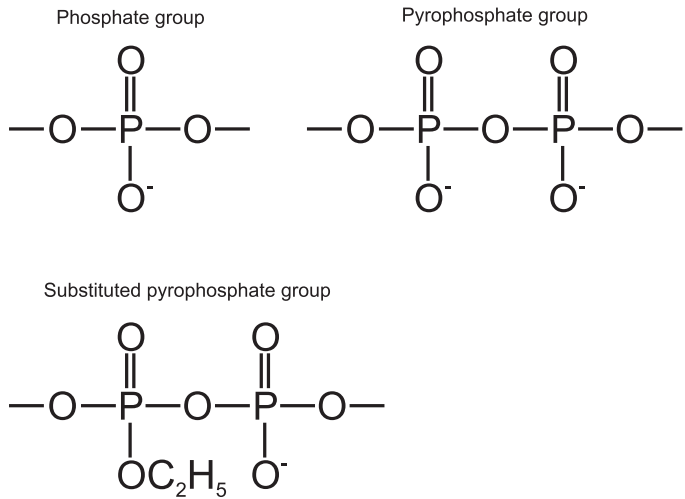}
    \caption{Phosphate, pyrophosphate and substituted pyrophosphate group.}
    \label{fig:p-groups}
\end{figure}

\begin{figure}[htb]
    \centering
        \includegraphics[width=0.5\textwidth]{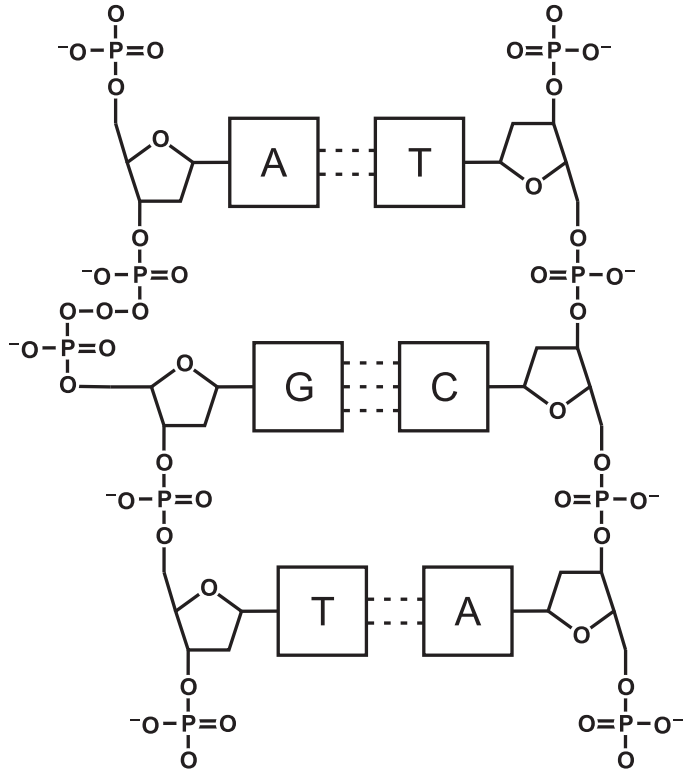}
   \caption{DNA with a pyrophosphate group.}
    \label{fig:p-dna}
\end{figure}

\section{\label{sec: Main} Model.}

Theoretical study of liquid crystalline phases of the DNA generally uses models that are necessarily
based on very crude simplifications.  The first point at issue is the right choice of a potential of
interaction. In this paper we model the molecular of the DNA on  a 1-dimensional lattice of charges
and dipoles with an elementary cell of size $3.4$ \AA. It mimics the spatial conformation of
charges of phosphate groups and dipoles of base-pairs. We consider short segments of the DNA,
approximately $500$ \AA, that is of the size of persistence length, so that to a good approximation
they are segments of straight lines, and assume that both molecules have the same number, 151,  of
base-pairs that can be visualized as points on a straight line parallel to the axis of the molecule,
one base-pair being located at the center of a corresponding molecule (see Figure 3).  The centers of
the straight lines belong to a straight line perpendicular to plane x-y which is  parallel to either
of them. We shall denote by $\xi$ the angle between the straight lines describing the molecules.  Both
molecules are of the same helicity, which is determined by the rotation of the frame of the dipole
moments. Thus, we model a molecule of the DNA on a one-dimensional lattice having at its sites either
vectors of dipoles of the base pairs or scalars of the phosphate charges. It is important that the
values of the dipoles and charges are renormalized owing to screening effects caused by counterions
and ions adsorbed at the molecule.  Therefore, we consider effective phosphate charges and dipoles of
base pairs. The case of total neutralization of phosphate charges was considered in paper \cite{kik}.

The dipoles are suggested to have the helix symmetry with  $\pi / 5$ rotation / bp, corresponding to
the structure of the ideal double helix of the DNA.   Of course, it is necessary to take into account
the structure of DNA being not uniform and the relative positions of the base pairs varying slightly
from  base pair to base pair.  Hence,  the dipole moments of the base pairs do not form a precise
lattice structure.  Even more so they should depend on the local DNA sequence.  Therefore, our
assumption of the regular dipole positions is a crude {\it approximation}.

The distance, $\kappa$, between the centers of the lattices, which is fixed, is an important parameter
of the model. In what follows we use the distance between adjacent base-pairs, that is $3.4$  \AA,
as a unit of length, take a unit of charge for which the dipole moment of $1 \; Debye$ equals $1$, and
perform calculations in the dimensionless units generated by these quantities.

The energy of electrostatic interaction of two molecules can be cast in the sum
\begin{equation}
    \label{U}
     \epsilon \, U = U_0 + u_{dd} + u_{dc} + u_{cd} + u_{cc}
\end{equation}
in which $\epsilon$ is the dielectric permeability of solvent and $U_0$ is the self energy of the
pair, which does not influence its conformation, the first term describes the interaction between
dipoles of the first molecule and those of the second; the second term - dipoles of the first and
phosphate charges of the second; the third - charges of the first and dipoles of the second; the
fourth - charges of the first and the second. The interactions are given by the equations
\begin{eqnarray}
    u_{dd}(\rho) &=& e^{-\nu \,\rho} \,
                     \left[ g(\rho) \frac{1}{\rho^3}
                             (\vec p \cdot \vec p^{\; \prime})    - 3 h(\rho)
                             \frac{ [\vec p \cdot (\vec r - \vec r^{\; \prime})]
                                    [\vec p^{\; \prime}
                                    \cdot (\vec r - \vec r^{\; \prime})]
                                  }
                                  {\rho^5}
                     \right]   \label{Udd}   \\
    u_{dc}(\vec r, \vec r^{\; \prime}) &=&  e^{-\nu \,\rho}  k(\rho)  Q^{\prime}
                                                       \frac{\vec p \cdot \vec r^{\; \prime}}{\rho}
                                                             \label{Ucd}\\
    u_{cd} (\vec r^{\; \prime}, \vec r) &=&  e^{-\nu \,\rho}   k(\rho)  Q
                                                       \frac{\vec p^{\; \prime} \cdot \vec r}{\rho}
                                                             \label{Udc}\\
    u_{cc}(\vec r, \vec r^{\; \prime}) &=&  e^{-\nu \,\rho}   \frac{ Q Q^{\prime}}{\rho}
                                                             \label{Ucc}
\end{eqnarray}
in which $\nu$ is the inverse Debye length $\nu = \lambda^{-1}$, and
$$
    \rho = |\vec r - \vec r^{\; \prime}|
$$
We shall take the screening functions $k(\rho), g(\rho), h(\rho)$ in Schwinger's form
$$
     k = g = 1 + \nu \, \rho, \quad h = 1 + \nu \, \rho + \frac{1}{3} \, \nu^2 \rho^2
$$

The important point about the electro-statical interaction between molecules of the DNA is  a wise
choice of the screening factor. The common practice  is to employ  the Debye-H\"uckel theory, or its
modifications that might accommodate the dipole charges, the so-called Schwinger screening,
\cite{podg}. The full treatment of this problem requires a separate investigation. In this paper we
confine ourselves to the Debye-H\"uckel and the Schwinger theories, \cite{podg}.

It is worth noting that the pair potential $U$ is invariant: if we change the sign of angle $\xi$
between the axes of the two molecules,   at the same time as the sign of helicity, the potential $U$
remains the same. There are  symmetry rules for the helixes of the same kind. One may convince oneself
that the following transformations
\begin{eqnarray}
     \label{t1}
        t_1 : \;
        (\phi_1, \; \phi_2, \; \xi) & \rightarrow & (\phi_1, \; \pi - \phi_2, \; \xi + \pi)  \\
    \label{t2}
        t_2 : \;
        (\phi_1, \; \phi_2, \; \xi) & \rightarrow & ( \pi - \phi_1, \; \phi_2, \; \xi + \pi)  \\
    \label{t3}
        t_3 : \;
        (\phi_1, \; \phi_2, \; \xi) & \rightarrow & (\phi_2 + \pi, \; \phi_1 + \pi, \; \xi)
\end{eqnarray}
leave the potential $U$ invariant. The angles are defined within limits
$$
     - \pi \le \phi_1 \le \pi, \quad
     - \pi \le \phi_2 \le  \pi, \quad
     - \pi \le \xi \le \pi
$$
values  $\pm \pi$  corresponding to the same configurations of the molecules. The transformations
given by equations (\ref{t1}-\ref{t3}) verify the equations
$$
    t^2_1 = t^2_2 = t^2_3 = id, \quad t_2 \, t_3 = t_3 \, t_1, \quad t_1 \, t_2 = t_2 \, t_1,
$$
where $id$ is a transformation that leaves all $\phi_1, \phi_2, \xi$ invariant. Using the above
equations one can easily convince oneself that $t_1, t_2, t_3$ generate a {\it non-commutative group}
of 8-th order, ${\cal S}$. Its maximal subgroup ${\cal H}$ is a normal subgroup of 4-th order,
commutative, and generated  by the transformations
\begin{equation}
   \label{f12}
      f_1 = t_3, \quad f_2 = t_1 \, t_2 \, t_3
\end{equation}
Elements $f_1, \; f_2$  in its turn generate subgroups ${\cal H}_1$ and ${\cal H}_2$ of ${\cal H}$,
respectively. It is worth noting that ${\cal H}_1, \;{\cal H}_2$ are of second order, both. They are
conjugate subgroups of {\cal S}, that is for an element $g$ of ${\cal S}$ we have $f_1 = g^{-1} \, f_2
\, g$, or we may state ${\cal H}_1 = g^{-1} \, {\cal H}_2 \, g$, in the notations of group theory,
which can be cast in the form of the diagram
\begin{equation}
    \label{conjug}
        {\cal H}_1 \longleftrightarrow {\cal H}_2
\end{equation}
The element
\begin{equation}
    \label{f3}
        f_3 = t_1 \, t_2
\end{equation}
generates subgroup ${\cal H}_3$ of ${\cal H}$. It is important that ${\cal H}_3$  is a normal subgroup
of ${\cal S}$, that is $g^{-1} \,  {\cal H}_3 \, g = {\cal H}_3 $ for any element $g$ of ${\cal S}$.
Thus, we have the diagram of subgroups inside the symmetry group ${\cal S}$
\begin{equation}
    \label{groupdiagram}
    \begin{array}{lllll}
        {\cal H}_1            &          &                 &                  \\
         & \; \searrow        &          &                 &                  \\
        {\cal H}_3 & \longrightarrow & {\cal H} & \longrightarrow & {\cal S}  \\
         & \; \nearrow        &          &                 &                  \\
        {\cal H}_2            &          &                 &
    \end{array}
\end{equation}
in which the arrows signify the embedding of subgroups.

The group of symmetries, ${\cal S}$, plays the key role in finding minima of  the potential $U$. The
following general arguments, based on the theory of groups, are quite useful in this respect. Consider
a point $\mu$ of  space ${\cal X}$ of the angles $\phi_1, \, \phi_2, \, \xi$. Suppose that $\mu$ is a
minimum of $U$. Then points
$$
    \mu^{\prime} = g \cdot \mu,
$$
called the orbit of the point $\mu$ under the action of the group ${\cal S}$, are also minima of $U$.
The number of points $\mu^{\prime}$ of  the orbit can vary. In fact, let us consider all
transformations $g$ of ${\cal S}$ that leave $\mu$ invariant, that is $\mu^{\prime} = g \cdot \mu =
\mu$. It is alleged to be known that the transformations form a subgroup of ${\cal S}$, called
stationary subgroup ${\cal H}_{\mu}$. The stationary subgroups, ${\cal H}_{\mu}$ and ${\cal H}_{\nu}$
, for points $\mu$ and $\nu$ of an orbit, are conjugate, that is $ {\cal H}_{\mu} = g^{-1}\, {\cal
H}_{\nu} \, g$ for an element $g$ of ${\cal S}$. The number of different points $\mu^{\prime}$ equals
to the ratio of the orders of ${\cal S}$ and ${\cal H}_{\mu}$, that is to 2 or 4, depending on the
choice of point $\mu$. To be specific, consider a point $\mu$ having a stationary subgroup ${\cal
H}_{\mu}$ that coincides with the subgroup ${\cal H}$. The latter is a normal subgroup of ${\cal S}$
of index 2, that is the factor set ${\cal S} /{\cal H}$ consists of two elements. Thus, the orbit of
$\mu$ under the action of ${\cal S}$ consists of only two points that correspond to the same value of
$U$ and have the same stationary subgroup ${\cal H}$, because the latter is a normal subgroup of
${\cal S}$. The situation is quite different if we take a point $\nu$ having stationary subgroup
${\cal H}_1$, which is different from ${\cal H}_2.$. The subgroups do not coincide in ${\cal S}$, even
though they are conjugate. The orbit of ${\cal \mu}$ under the action of ${\cal S}$ indicated above
consists of four points that we may sort out as follows: two points having the stationary subgroup
${\cal H}_1$ and two points having ${\cal H}_2$.  This is due to the fact that for one thing the
subgroup ${\cal H}$ is commutative and therefore its elements generate points of the orbit but with
the same stationary subgroup, that is ${\cal H}_1$, and for another there is an element $g$ that gives
points of the orbit having the stationary subgroup ${\cal H}_2$.  In contrast, a point $\mu$ having
the stationary subgroup ${\cal H}_3$ has the orbit consisting of four points which have the same
stationary subgroup ${\cal H}_3$, because the latter is a normal subgroup of ${\cal
S}$.\label{statsubgr}

\section{Numerical simulation}

\begin{figure}[htb]
    \centering
        \includegraphics[width=0.4\textwidth]{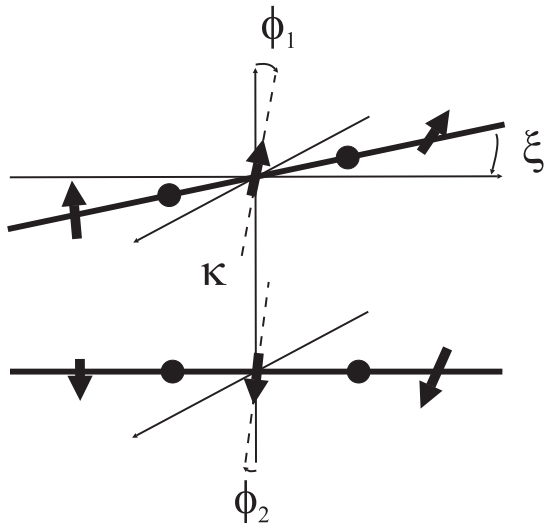}
    \caption{
       Scheme for the locations of dipoles and charges of base pairs and P-groups, respectfully,
       in a molecule of the DNA. The conformation of a pair of molecules is determined by angle $\xi$
       between the axes of the molecules, and angles $\phi_1, \; \phi_2$ of rotations of the molecules
       about their axes.
    }
    \label{fig:model}
\end{figure}

It is to be noted that numerical evaluation of the minima runs across a poor convergence of standard
algorithms for minimization, owing to flat  surfaces of constant value for the function of three
variables, $U(\phi_1, \; \phi_2, \; \xi)$. To some extent, one may get round the difficulty by
observing that for points remaining fixed with respect to a subgroup ${\cal G}$ of ${\cal S}$, the
minimization problem is reduced to that for a smaller number of variables. This is due to the
necessary conditions for extremum being verified automatically for degrees of freedom perpendicular to
the set of invariant points, so that one needs only to study the conditions for longitudinal
variables, that is to solve a smaller system of equations.

To see the point let us consider a function $f(x,y,z)$ of variables $x, y, z$ even in $x$, so that
$f(x,y,z) = f(- \, x, y,z)$. The set of invariant points is $y-z$ plane, and we may look for minima of
the function $f(x=0,y,z)$, thus we need to solve  only  two equations
$$
    \frac{\partial}{\partial y}f(x=0, y,z) = 0, \quad \frac{\partial }{\partial z}f(x=0, y,z) = 0
$$
The number of variables necessary for calculations can be reduced even further in case of larger
groups of symmetries. It is easy to convince oneself that the sets of fixed points $(\phi_1, \phi_2,
\xi)$, that is invariant under the action of a subgroup of ${\cal S}$, read as follows
\begin{eqnarray}
    {\cal F}_1 &:&  (\phi_1 = \phi, \; \phi_2 = \phi + \pi, \; \xi) \label{F1}   \\
    {\cal F}_2 &:&  (\phi_1 = \phi, \; \phi_2 = - \, \phi , \; \xi) \label{F2} \\
    {\cal F}_3 &:&  (\phi_1 = \pm \, \frac{\pi}{2}, \; \phi_2 = \pm \,\frac{\pi}{2} , \; \xi)
    \label{F3}
\end{eqnarray}
in which the ${\cal F}_i$ are invariant under the action of subgroups ${\cal H}_1, \; {\cal H}_2, \;
{\cal H}_3$, respectively. The above symmetries are illustrated in Figure 4.

 \begin{figure}[htb]
    \centering
     \includegraphics[width=1.\textwidth]{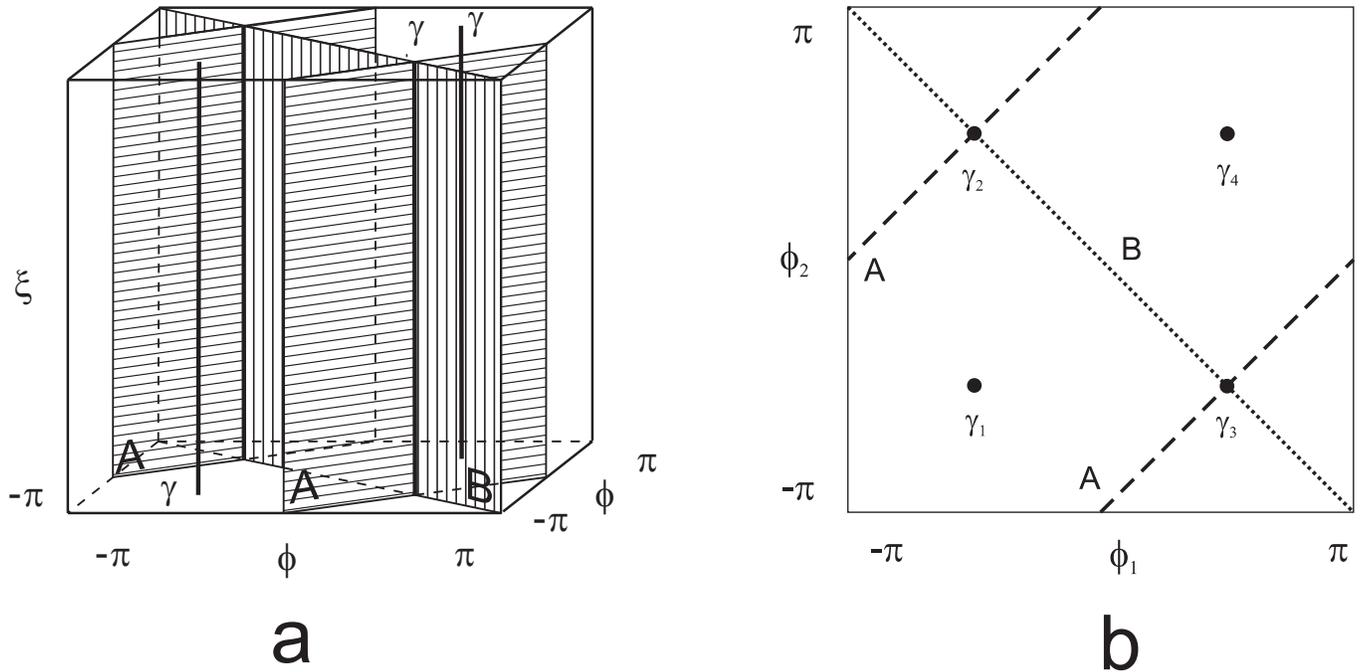}
     \caption{
          \label{symmetrygraph}
          (a) Cube of the symmetries indicating sets in space $\phi_1, \phi_2, \xi$
          invariant with respect to  subgroups of ${\cal S}$.
          Main diagonal plane, $B$, corresponding to subgroup ${\cal H}_2$;
          two rectangles perpendicular $A$ to ${\cal H}_1$;
          solid lines $\gamma_1, \; \gamma_4$, and $\gamma_2, \; \gamma_3$
          corresponding, to ${\cal H}_3$ and ${\cal H}$, respectfully.
          (b) Cube of the symmetries view from the top.
          Dotted line describes invariant points ${\cal F}_2$, corresponding to
          subgroup ${\cal H}_2$; dashed line points ${\cal F}_1$ and
          subgroup ${\cal H}_1$;
          solid circles $\gamma_1$ and $\gamma_4$  to subgroup ${\cal H}_3$;
          $\gamma_2$ and $\gamma_3$  to subgroup ${\cal H}$.
     }
 \end{figure}

The analysis of symmetries of  $U$ given above, cf. p.~\pageref{statsubgr}, enables us to sort out the
minima according to the effective value  of the  phosphate charge $Q$. The dependence of the values of
minima on the effective charge is illustrated in Figure 5.

\begin{figure}[htb]
    \centering
        \includegraphics[width=0.6\textwidth]{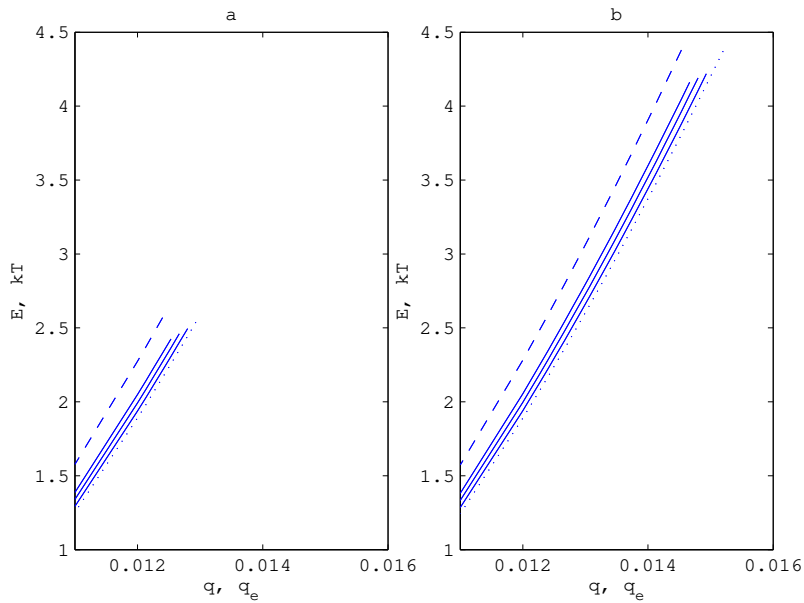}
    \caption
    {
        Minima of $U$ in units of $k_B T, \; T = 300 K$:
        (a) cholesteric angle $\xi > 0$;
        (b) $\xi < 0$;
        against effective charge $q$ in units of the electron charge.
        Values of $U(\xi, \phi_1, \phi_2)$ and $U(- \; \xi, \phi_1, \phi_2)$ are equal to within $0.01 \; k_B
        T$.
        Minima that could correspond to cholesteric phase with $\xi > 0$ vanish at $q = 0.013$
    }
    \label{fig:chargeagainstenergy-separated}
\end{figure}

 \begin{figure}[htb]
    \centering
     \includegraphics[width=0.6\textwidth]{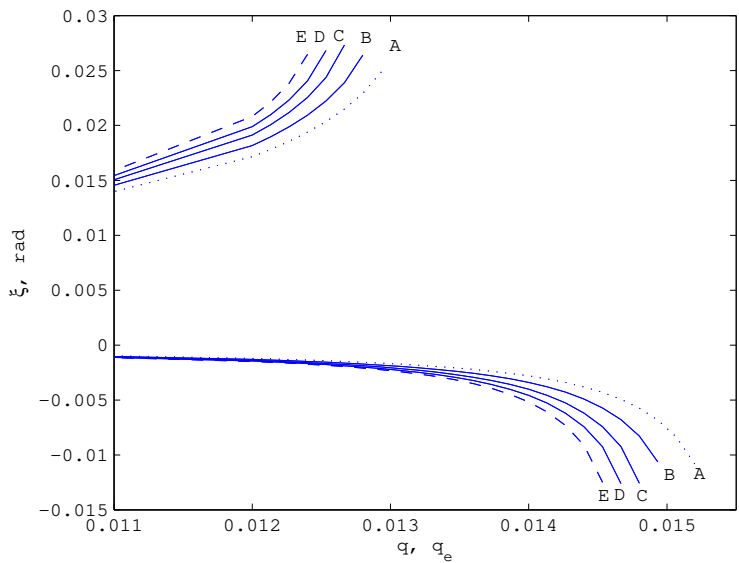}
     \caption{\label{q_E}
         Cholesteric angle $\xi$ against effective charge $q$ in units of the electron charge.
         PP-groups are located symmetrically:
         (A) no PP-groups;
         (B) one PP-group at either end of a molecule;
         (C) two PP-groups at either end of a molecule;
         (D) three PP-groups;
         (E) every tenth P-group is exchanged for the PP-one.
    }
 \end{figure}

It should be noted that minima of the pair-interaction $U$ depend on the distance between molecules
$\kappa$, and the  effective phosphate charge $Q$. The latter is the control parameter we employ in
numerical simulation. It is also useful for the description of possible experimental results. In this
paper we are considering $\kappa$ to within $10.2 \pm 34$ \AA. Effective charge $Q$ of phosphate
groups determines the neutralization; it varies to within $0 \pm .6$, in dimensionless units, $Q=0$
corresponding to the total neutralization. Charges $Q$ that correspond to the charge inversion, have
not been considered.  The Debye length, $\lambda$, has been varied to within $7 \pm 35$ \AA,
depending on the ion strength of solution.

The value of effective charge $Q$ is determined by its electrostatic surrounding.  It depends not only
on ion charges in solvent, but also on those adsorbed by a molecule of the DNA.It seems that the
conventional Debye --- H\"uckel theory does not work in this situation, \cite{podg}. At any rate, it
does not accommodate the adsorbed charges.According to \cite{livolant2} the effective charge is small.

The numerical data and the symmetry analysis given above suggest that there should be the following
three types of the minima. \vspace{-3mm}
\begin{enumerate}
    \item Type I characterized by  molecules having a
          cross-like conformation, " snowflakes", that is $\xi$ being  close to $\pi /2$.
          It exists for $Q$ large enough.
          Its symmetry subgroup depends on  $Q$ and
          may take values ${\cal H}_1, \; {\cal H}_2, \; {\cal H}_3, \;{\cal H}$.
          Therefore, we may claim that there exist four sub-types of minima I:
          $\rm{I}_{{\cal H}_1}, \; I_{{\cal H}_2}, \;I_{{\cal H}_3}, \;I_{{\cal H}}$,
          each of them consisting of two subtypes which are given by specific conformations of
          the angle variables.
          \vspace{-5mm}\\
    \item Type II for which $\xi$ takes  values to within $0.1^{o}$;
          the symmetry subgroups are ${\cal H}_1$ and ${\cal
          H}_2$, either type consists of two sub-types.
    \item Type III for which $\xi$ is to within
          $1^{o}$, that is larger than for II. The symmetry
          subgroup is ${\cal H}_3$, and there are four constituent types of the same symmetry.
 \end{enumerate}

As can be inferred from the considerations given above,  the study of the pair-interactions between
molecules of the DNA requires a means to vary the charges of a molecule and their positions in it. By
now the only method available to that end is to vary the ion composition of solvent, the molecules
itself being intact. The use of the DNA  containing PP-groups, should provide new opportunities for
the research, for it could change in a prescribed way  the conformation of charges of the molecule.
Thus one may compare the formation of liquid crystalline phases for the same solutions but for a
different charge conformation of the DNA.  Our numerical simulation suggests that the effect could be
tangible enough to be looked in experiment. The dependence of the minima on effective charge taking
into account  PP-groups is indicated in Figure 6. The behavior of $U$ is illustrated in Figure 7 by
means of iso-energy surfaces. Another point in favor of working with the pyro-phosphate forms of the
DNA is that one can vary the effective charges of a molecule. In the case of the usual phosphate DNA
the phosphate charges are all equal, and therefore one may suggest that the effective charges, which
enter in our simulation, are also equal. Using the pyro-phosphate forms we may expect to achieve even
the regime for which all the phosphate charges are neutralized whereas the pyro-phosphate ones still
remain, even though being small.  Such an experiment would be helpful in determining the nature of
intra-molecular pair interaction that leads, according to the accepted physical picture,
\cite{lk_model}, to the formation of cholesteric phases of the DNA.

The main point about  our numerical simulation is the choice of  values for effective charges and
dipoles.  In case there are PP-groups we may consider the effective charges in a way described above.
The situation is more subtle as far as dipoles are concerned. We assume their  numerical values being
of the first order in the units indicated above.   This is as much as to say that the charges of base
pairs that constitute dipoles, are screened much less than the phosphate ones.  If we proceed
otherwise and take small values of the dipoles, there are no minima small but nonzero value of $\xi$,
that is no cholesteric phases. In fact, they are, \cite{livolant1}.  Thus,  it is reasonable to assume
that the charges of base pairs are screened in a manner different from that for the phosphate ones.
One may suggest to the effect that the renormalization of charges is due mainly to adsorption of ions
from the solvent, and not to the screening clouds of ions in the solvent.  If so, it is likely that
the charges of P-groups, due to ions of $O^{-}$, are in a different positions than those of the base
pairs,  and  the renormalization of charges of P-groups and base pairs follow different paths.  For
this argument we are indebted to Yu.M. Yevdokimov.

The important point is that PP-groups make for the formation of cholesteric angles different from
those of P-groups, and thus provide a means of the identification of new phases.  Summing up:
\begin{itemize}
    \item the  phase of "snowflakes" sustains the presence of PP-groups;
    \item the PP-groups may result in splitting energy levels of minima,
          so that minima  corresponding to the same values of $U$ become
          different, when PP-groups are present;
    \item minima of $U$ are  separated by low potential barriers; iso-energy surfaces
          of constant values of $U$,  being like galleries between halls that illustrate the minima;
    \item minima corresponding to cholesteric phases are very narrow,
          whereas those of snowflakes are  broad and sloping; energy barriers separating minima
          corresponding to snowflakes and cholesteric phases, respectively, are of the order $k_{B}T$;
    \item values of angles $\phi_1, \; \phi_2$ are subject to constraints  inside  galleries joining two minima,
\end{itemize}

\begin{figure}[htb]
    \centering
        \includegraphics[width=1.\textwidth]{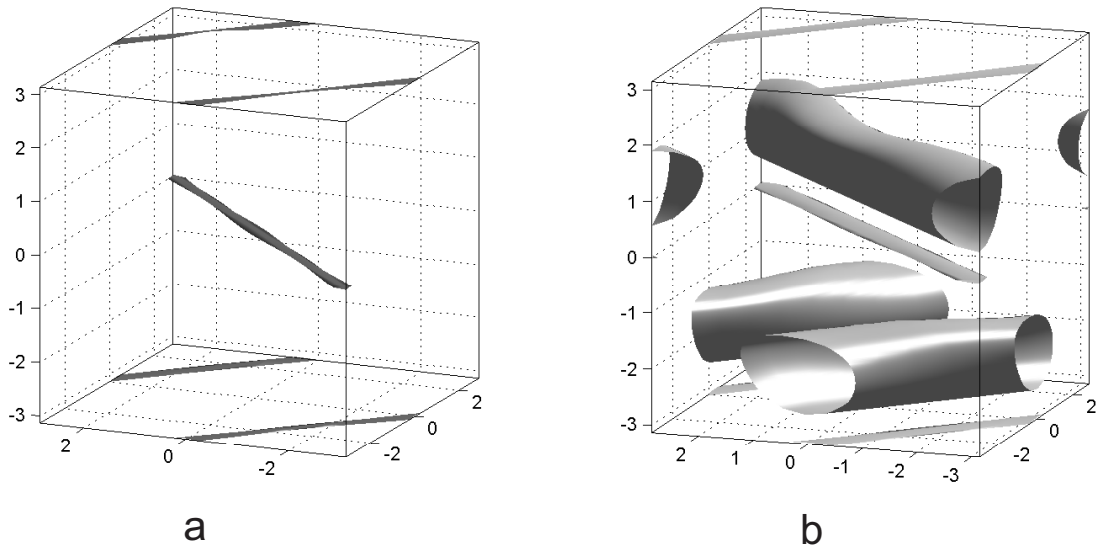}
   \caption{ Surfaces of constant value of $U$  near  cholesteric  minima.
             (a) Screened charge $q = 0.0086$ $q_e$, energy level is $0.05 \; kT$.
             (b) Screened charge $q = 0.0086$ $q_e$, energy level is $0.2 \;kT$.
    \label{fig:cube1}
   }
\end{figure}

\section{Conclusions: New opportunities for studying the liquid crystalline phases of the DNA.  }

The pyro-phosphate DNA may serve a valuable probe into the physics of cholesteric phases of the DNA.
providing a unique opportunity for changing the effective charge of a molecules of the DNA. The use of
PP-forms  may result in appreciable experimental effects, which in its turn could throw light on the
nature of intramolecular interaction in solution of the DNA. The charge screening still poses a number
of questions.  The usual Debye - H\"uckel theory does not seem to be an adequate solution,
\cite{podg}, especially as the screening of electrical dipole moments is concerned. A theory that
could give a reasonable agreement with experiment, should be  that of finite number of particles,
whereas  the Debye - H\"uckel theory relies on the use of macroscopical considerations.  The study of
the cholesteric phases of the DNA with pyrophosphate inter-nucleotide insertions could provide a means
to find  characteristics that indicate a way to understanding the phenomenon. An important issue is
the different screening of the phosphate charges and the dipoles of the base pairs.  It is  should be
noted that the screening is caused by a non-uniform adsorption of counterions at a molecule of the
DNA, so  that  the phosphate charges and the base pair dipoles are not screened in the same manner.

Our calculations rely on a model that is based on  general and qualitative assumptions of the helical
charge distribution of the DNA.  We feel that it accommodates a picture of the DNA, without going into
finer details, and agrees  with the approach of paper \cite{lk_model} in which the continuous
approximation plays the directive part.  The choice of the pair-potential for the intramolecular
interaction is important. The shape of the pair potential chosen in this paper enabled us to
accommodate the two different sets of charges --- the Coulomb and the dipole ones, and also take into
account  a finer detail of the pyrophosphate charges, which could turn out to be a valuable instrument
for further investigating the cholesteric phases. The symmetry constraints have played an important
part in finding the minima of the pair-potential. It seems that their meaning could be greater than a
simple arithmetic device for simplifying calculations, and  could indicate certain symmetry law
peculiar to the cholesteric phases of the DNA. It is reasonable to expect the polymorphism  liquid
crystalline phases of the DNA.  New artificially synthesized phases of the DNA could be a fruitful
instrument to that end.

We are thankful to F.Livolant for the useful correspondence. This work was done within the framework
of the Interdisciplinary Programme of the Lomonosov Moscow State University, and partially supported
by RFBR  Grant \# NS---660.2008.1.

\end{document}